\documentclass[a4paper,10pt,twoside]{cpc-hepnp}
\usepackage{multicol}
\usepackage{graphicx}
\usepackage{booktabs}
\usepackage{amssymb,bm,mathrsfs,bbm,amscd}
\usepackage{lastpage}
\usepackage{changepage} 

\usepackage[scientific-notation=true]{siunitx}
\usepackage{multirow}
\usepackage{tikz}
\usetikzlibrary{shapes.geometric, arrows}

\usepackage[utf8]{inputenc}
\usepackage[colorlinks=true, linkcolor=red, citecolor=blue]{hyperref}

\usepackage{amsmath,amsfonts,amssymb,bm}
\usepackage{graphicx}
\usepackage{color}
\usepackage{subfigure}
\usepackage{multirow}
\usepackage{textcomp}
\usepackage{slashed}
\usepackage{cancel}

\def\({\left(}
\def\){\right)}
\def\[{\left[}
\def\]{\right]}

\def\be{\begin{eqnarray}}
\def\ee{\end{eqnarray}}
\usepackage[capitalise]{cleveref}
\crefname{figure}{Fig.}{Figs.}
\Crefname{figure}{Fig.}{Figs.}

\begin{document}

\fancyhead[c]{\small Submitted to Chinese Physics C} \fancyfoot[C]{\small Page-\thepage}



\title{Scalar Mesons and Axial-vector Mesons Via Dyson-Schwinger Equation and Bethe-Salpeter Equation Approach  \footnote{This work is supported by: the National Natural Science Foundation of China under contracts No. 11947108 and No. 12005060.}}

\author{%
      Muyang Chen $^{1}$\email{muyang@hunnu.edu.cn}
}

\maketitle

\address{%
$^1$ Department of Physics, Hunan Normal University, Changsha 410081, China
}

\begin{abstract}
The spectrum of scalar mesons and axial-vector mesons are systematically studied via Dyson-Schwinger equation and Bethe-Salpeter equation approach in the rainbow-ladder approximation. An interaction model with a repulsive term added to the one used for the pseudoscalar and vector mesons is proposed. The Dyson-Schwinger equation results are consistent with the experiment values and other model results, which shows that this interaction model is effective for all the heavy, heavy-light, and light scalar and axial-vector mesons. This is the first time a systematic study of the spectrum of the scalar mesons and axial-vector mesons in the Dyson-Schwinger equation approach is achieved.
\end{abstract}




\begin{multicols}{2}

\section{Introduction}
\label{sec:intro} 
\noindent

A systematic study of the hadrons in a Quantum Chromdynamics (QCD) related way is important for the hadron phenomenon. The Dyson-Schwinger equations (DSEs) \cite{Dyson1949,Schwinger1951} provide such a way. This approach preserves important properties of QCD such as dynamical chiral symmetry breaking (DCSB), and many phenomenological successfulness have been gained \cite{Eichmann2016,Roberts2017}.

Studying the meson spectrum via Dyson-Schwinger equation and Bethe-Salpeter equation (DSBSE) approach \cite{Dyson1949,Schwinger1951,Salpeter1951} using realistic interaction begun 20 years ago. A gaussian interaction model in the rainbow ladder (RL) approximation is quite successful for the light pseudoscalar and vector mesons \cite{Maris1997,Maris1998,Maris1999,Qin2011}. This model was extended to study the heavy mesons \cite{Fischer2015,Hilger2015} and heavy-light mesons \cite{Chen2019} in recent years. Combining with the Faddeev equation, this approach can also study the baryons \cite{Eichmann2016,Qin2020}. 

The DSBSE approach can be briefly stated as following. The quark propagator is solved by the Gap equation \cite{Dyson1949,Schwinger1951,Roberts2017}  \footnote[2]{We work in the Euclidean space, where the inner product of the four vector is defined by $a\cdot b = \delta_{\mu\nu}a_\mu b_\nu = \Sigma_{i=1}^4  a_i b_i$, with $\delta_{\mu\nu}$ being the Kronecker delta. The Dirac matrices satisfy the algebra $\{\gamma_\mu,\gamma_\nu \} = 2\delta_{\mu\nu}$, and $\gamma_5=-\gamma_1\gamma_2\gamma_3\gamma_4$.},
\begin{eqnarray}\nonumber
 S_f^{-1}(k) &=& Z_2 (i\gamma\cdot k + Z_m m_f) \\\label{eq:DSE0}
&&+ \frac{4}{3}\bar{g}^2 Z_1 \int^\Lambda_{d q} D_{\mu\nu}(l)\gamma_\mu S_{f}(q)\Gamma^f_\nu(k,q),
\end{eqnarray}
where $f=\{u,d,s,c,b,t\}$ represents the quark flavor. $S_{f}(k)$ is the quark propagator, which can be decomposed as $S_{f}(k) = \frac{Z_f(k^2)}{i\slashed{k} + M_f(k^2)}$. $Z_f(k^2)$ is the quark dressing function and $M_f(k^2)$ is the quark mass function. $l=k-q$. $m_f$ is the current quark mass. $\Gamma^f_\nu$ is the quark-gluon-vertex. $D_{\mu\nu}$ is the gluon propagator. $\bar{g}$ is the coupling constant. $Z_1$, $Z_2$, $Z_m$ are the renormalisation constants of the quark-gluon-vertex, the quark field and the quark mass respectively.
$\int^\Lambda_{d q}=\int ^{\Lambda} d^{4} q/(2\pi)^{4}$ stands for a Poincar$\acute{\text{e}}$ invariant regularized integration, and $\Lambda$ is the regularization scale.

The meson Bethe-Salpeter amplitude (BSA), $\Gamma^{fg}(k;P)$, is solved by the Bethe-Salpeter equation (BSE). $k$ and $P$ are the relative and total momentum of the meson. $P^2 = -M^2_{fg}$ and $M_{fg}$ is the mass of the meson with valence quark flavor $(f,g)$. The BSE expresses as \cite{Salpeter1951,Roberts2017}
\begin{equation}\label{eq:BSE0}
  \big{[} \Gamma^{fg}(k;P)  \big{]}^{\alpha}_{\beta}  =   \int^\Lambda_{d q} \big{[} K^{fg}(k,q;P) \big{]}^{\alpha\delta}_{\sigma\beta} \big{[} \chi^{fg}(q;P)  \big{]}^{\sigma}_{\delta} ,
\end{equation}
where $ K^{fg}(k,q;P)$ is the quark-antiquark scattering kernel, and $\alpha$, $\beta$, $\sigma$ and $\delta$ are the Dirac indexes.
$\chi^{fg}(q;P) = S_{f}(q_{+}) \Gamma^{fg}(q;P) S_{g}(q_{-})$ is the wave function.
$q_{+} = q + \eta P/2$, $q_{-} = q - (1-\eta) P/2$. $\eta$ is the partitioning parameter describing the momentum partition between the quark and antiquark and varying its value dosen't affect the physical observables.

The RL approximation is replacing the self energy in Eq. (\ref{eq:DSE0}) and the scattering kernel in Eq. (\ref{eq:BSE0}) by
\begin{eqnarray}\label{eq:quarkRL}
&& \bar{g}^2 Z_1 D_{\mu\nu}(l) \Gamma^f_\nu(k,q) \to [Z_{2}]^{2} \tilde{D}^{f}_{\mu\nu}(l) \gamma_\nu, \\\label{eq:mesonRL}
&& \big{[} K^{fg}(k,q;P) \big{]}^{\alpha\delta}_{\sigma\beta} \to -\frac{4}{3}[Z_{2}]^{2} \tilde{D}^{fg}_{\mu\nu}(l) [\gamma_{\mu}^{}]^{\alpha}_\sigma [\gamma_{\nu}]^\delta_\beta,
\end{eqnarray}
where $\tilde{D}^{f}_{\mu\nu}(l) = \left(\delta_{\mu\nu}-\frac{l_{\mu}l_{\nu}}{l^{2}}\right)\mathcal{G}^f(l^2)$ and $\tilde{D}^{fg}_{\mu\nu}(l) = \left(\delta_{\mu\nu}-\frac{l_{\mu}l_{\nu}}{l^{2}}\right)\mathcal{G}^{fg}(l^2)$ are the flavor dependent effective interaction between the quark and antiquark. In Ref. \cite{Chen2019}
the dressing function $\mathcal{G}^f(l^2)$ and $\mathcal{G}^{fg}(l^2)$ are modeled as
\begin{eqnarray}\label{eq:gluonmodel}
  \mathcal{G}^f(s) 	&=& \mathcal{G}^f_{IR}(s) + \mathcal{G}_{UV}(s),\\\label{eq:gluonInfrared}
  \mathcal{G}^f_{IR}(s) &=& 8\pi^2\frac{D_f^2}{\omega_f^4} e^{-s/\omega_f^2},\\\label{eq:gluonfmodel}
  \mathcal{G}^{fg}(s) 	&=& \mathcal{G}^{fg}_{IR}(s) + \mathcal{G}_{UV}(s),\\\label{eq:gluonfInfrared}
  \mathcal{G}^{fg}_{IR}(s) 	&=& 8\pi^2\frac{D_f}{\omega_f^2}\frac{D_g}{\omega_g^2} e^{-s/(\omega_f\omega_g)},\\\label{eq:gluonUltraviolet}
  \mathcal{G}_{UV}(s) 	&=& \frac{8\pi^{2} \gamma_{m}^{} \mathcal{F}(s)}{\text{ln}[\tau+(1+s/\Lambda^{2}_{QCD})^2]},
\end{eqnarray}
where $s=l^2$. $\mathcal{G}^f_{IR}(s)$ and $\mathcal{G}^{fg}_{IR}(s)$ are the infrared interaction responsible for DCSB. $1/\omega_f$ is the interaction width and $D_f$ expresses the infrared interaction strength. $\mathcal{G}_{UV}(s)$ keeps the one-loop perturbative QCD limit in the ultraviolet.
$\mathcal{F}(s)=[1 - \exp(-s^2/[4m_{t}^{4}])]/s$, $\gamma_{m}^{}=12/(33-2N_{f})$, with $m_{t}=1.0 \textmd{ GeV}\,$, $\tau=e^{10} - 1$, $N_f=5$, and $\Lambda_{\text{QCD}}=0.21 \textmd{ GeV}\,$. This model turned out to be successful for all the ground state pseudoscalar and vector mesons, from heavy, heavy-light to light mass scale.

However, the orbital and radial excited states are less studied. The interaction models successful for pseudoscalar and vector mesons usually don't produce a good spectrum for the orbital and radial excited states \cite{Qin2011,Krassnigg2011}.
Even in the heavy quark region, the masses of the excited states are lower than the experiment data systematically \cite{Fischer2015,Hilger2015,Chen2017,Chen2020,Chang2020} because the repulsive interaction from higer order corrections of the quark-gluon vertex is lacking \cite{Bender1996}. One way to solve this problem is using beyond RL approximations. The effects of beyond RL approximations have been researched for years \cite{Chang2009,Qin2014,Binosi2016}. However, no systematic study of the hadron spectrum was achieved due to the technique difficulties. In this background, phenomenological method is of practical significance. In this work, I follow the latter way. The previous model, Eq. (\ref{eq:gluonmodel}) $\sim$ (\ref{eq:gluonUltraviolet}), is extended to study the scalar mesons and axial-vector mesons. In section \ref{sec:model}, I introduce the modified interaction model. In section \ref{sec:results}, I show the results and give some discussions. In section \ref{sec:summary}, I give a summary.

\section{The modified model}\label{sec:model}
\noindent

To make the interaction more reliable for scalar mesons and axial-vector mesons, Eq. (\ref{eq:gluonfInfrared}) is modified into
\begin{equation}\label{eq:gluonfdInfrared}
   \mathcal{G}^{fg}_{IR}(s) = 8\pi^2 (\frac{D_f}{\omega_f^2}\frac{D_g}{\omega_g^2} - \Delta_{fg}) e^{-s/(\omega_f\omega_g)},
\end{equation}
where $\Delta_{fg}>0$ represents a repulsive interaction that is missing in the RL approximation. An interaction universial for different mass scales should depend on the quark masses. The following form is proposed,
\begin{equation}\label{eq:gluondInfrared}
 \Delta_{fg} = \frac{\delta}{[M_f(0)M_g(0)]^{\alpha/2} (\omega_f\omega_g)^{\beta/2}},
\end{equation}
where $M_f(0)$ is the value of the quark mass function at $k^2=0$. $\Delta_{fg}$ in Eq. (\ref{eq:gluondInfrared}) depends on the quark masses and the interaction width explicitly. $\delta$, $\alpha$ and $\beta$ are 3 parameters to be fitted by the observables.

In this paper, I consider 5 quark flavors and work in the isospin symmetry limit. So $f,g = \{n,s,c,b\}$, with $n = u\, \textmd{or}\, d$. Herein I choose the masses of scalar ($J^{PC} = 0^{++}$) charmonium ($M_{c\bar{c}}$), bottomium ($M_{b\bar{b}}$) and the scalar ($J^{P} = 0^{+}$) D meson ($M_{n\bar{c}}$) to fit the three parameters. The masses of the other ground state scalar mesons and all the ground state axial-vector mesons are outputs of this model.

Before discussing the results, there is one more ingredient to be stated. The quark propagators in Eq.~(\ref{eq:BSE0}) are functions of the complex momenta $q^{2}_{\pm}$ which lie in parabolic regions. As the dressed quark propagators possess time-like complex conjugate mass poles \cite{Bhagwat2003}, there is an upper bound of the bound state mass obtainable directly, i.e. $P^2 > -M_{max}^2$, where $M_{max}^2$ defines the contour border of the parabolic region. See the appendix of Ref. \cite{Hilger2015} for this problem.

We adopt the extrapolation scheme to overcome this difficulty. This method has been used to study the radial excited $B_c$ mesons \cite{Chang2020} and a brief description is given below. Eq. (\ref{eq:BSE0}) is solved as a $P^{2}$-dependent eigenvalue problem,
\begin{equation}\label{eq:BSE2}
  \lambda^{fg}(P^2) \big{[} \Gamma^{fg}(k;P)  \big{]}^{\alpha}_{\beta}  =   \int^\Lambda_{d q} \big{[} K(k,q;P) \big{]}^{\alpha\delta}_{\sigma\beta} \big{[} \chi^{fg}(q;P)  \big{]}^{\sigma}_{\delta}.
\end{equation}
The meson mass is determined by $\lambda^{fg}(P^2=-M^2_{fg})=1$. If the meson mass is beyond the contour border, i.e. $M_{fg} > M_{max}$, then the following form is used to fit $\lambda^{fg}(P^2)$,
\begin{equation}\label{eq:fitlambda}
 \frac{1}{\lambda^{fg}(P^2)} = \frac{1 + \sum^{N_{o}}_{n=1}\, a_{n} (P^{2} + s_{0}^{})^n}{1 + \sum^{N_{o}}_{n=1}\, b_{n} (P^{2} + s_{0}^{})^n} \, .
\end{equation}
In Eq. (\ref{eq:fitlambda}) $N_{o}$ is the order of the series, $s_{0}^{}$, $a_{n}$ and $b_{n}$ are the parameters to be determined by the least square method.

\begin{center}
 \includegraphics[width=0.5\textwidth]{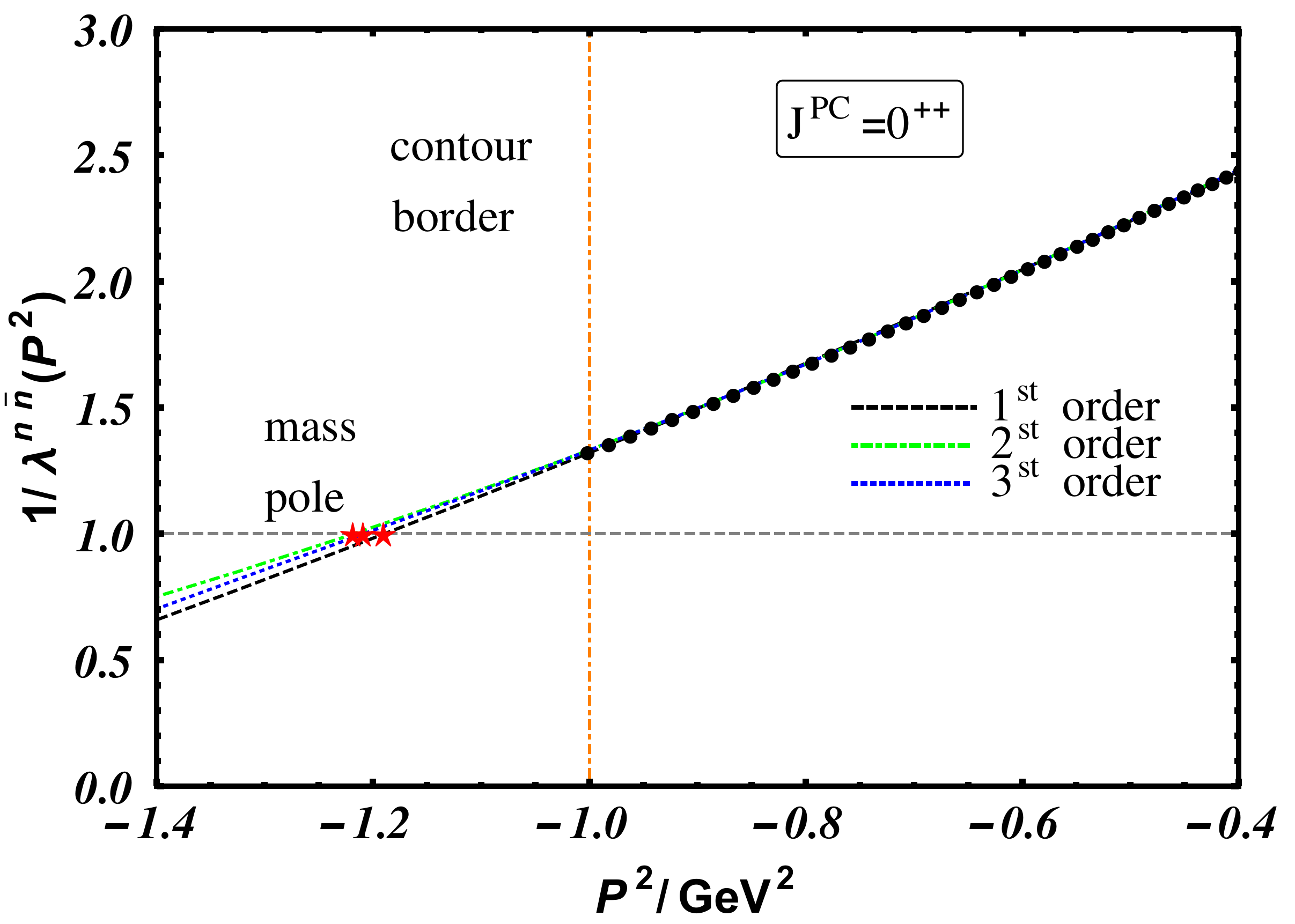}
\vspace*{-3mm}
\figcaption{(color online) The $P^{2}$ dependence of $1/\lambda^{n\bar{n}}$ for $J^{PC} = 0^{++}$ state using the middle group of parameters in Table. \ref{tab:parameters}. The vertical dot-dashed orange line is the contour border on the right of which the direct calculation can be applied. The black points are the calculated values. The dashed black line, dot-dashed green line and dotted blue line are the extrapolation functions, $1/\lambda^{n\bar{n}}(P^2)$, with $N_{o} = 1,\,2,\,3$ in Eq. (\ref{eq:fitlambda}) respectively. The red stars present the location of the meson masses.}
\label{fig:lambdaPP}
\end{center}

The extrapolation results are illustrated by the lightest scalar meson, shown in figure \ref{fig:lambdaPP}. As the mass pole is not too far away from the contour border, the series in Eq. (\ref{eq:fitlambda}) converge very well. So the error from extrapolation is controllable. In practice $N_o = 1, 2, 3$ is considered, and the maximal difference between them is estimated as the error from extrapolation.

\section{Results and Discussion}\label{sec:results}
\noindent

In Ref. \cite{Chen2019}, three group of parameters are used for $\{\bar{m}_f^{\zeta=2\textmd{GeV}}, \omega_f, D_f \}$, $f = \{n, s, c, b\}$, $n = u\, \textmd{or}\, d$. Herein I also consider these three cases. In each case, $\delta$, $\alpha$ and $\beta$ in Eq.(\ref{eq:gluondInfrared}) are fitted by three observables: the masses of scalar ($J^{PC} = 0^{++}$) charmonium ($M_{c\bar{c}}$), bottomium ($M_{b\bar{b}}$) and the scalar ($J^{P} = 0^{+}$) D meson ($M_{n\bar{c}}$). The parameters are listed in table \ref{tab:parameters} in the appendix. The masses of the other ground state scalar mesons and all the ground state axial-vector mesons are the outputs, listed in table \ref{tab:heavymesons}, \ref{tab:heavylightmesons} and \ref{tab:lightmesons}.
In these tables, two errors of the DSE results are given. The first error is from the extrapolation, the second one is from varying the parameters. Next, let's discuss the results of heavy mesons, heavy-light mesons and light mesons respectively.

\subsection{The heavy mesons}

The results of the heavy mesons are listed in table \ref{tab:heavymesons}. While the scalar ($J^{PC} = 0^{++}$) charmonium and bottomium masses are used to fit the parameters, the axial vector ($J^{PC} = 1^{++} \textmd{ or } 1^{+-}$) counterparts are the outputs. Comparing the DSE results with the experiment values, the largest deviation comes from the $J^{PC} = 1^{+-}$ charmonium, being about $20\textmd{ MeV}$. The other three mass deviations are within $10\textmd{ MeV}$. The scalar ($J^{P} = 0^{+}$) and lowest two axial-vector ($J^{P} = 1^{+} \textmd{ or } 1^{-}$) $B_c$ meson masses are also output quantities. Note that the DSE results from the modified model (Eq.(\ref{eq:gluonmodel}),(\ref{eq:gluonInfrared}),(\ref{eq:gluonfmodel}),(\ref{eq:gluonfdInfrared}),(\ref{eq:gluondInfrared}),(\ref{eq:gluonUltraviolet})) are almost the same as those from the original model (Eq.(\ref{eq:gluonmodel}),(\ref{eq:gluonInfrared}),(\ref{eq:gluonfmodel}),(\ref{eq:gluonfInfrared}),(\ref{eq:gluonUltraviolet})) after subtracting the errors carefully \cite{Chen2020}. As there is no experiment data for these $B_c$ meson masses yet, the DSE results are compared with the lattice QCD \cite{Mathur2018} and quark model predictions \cite{Li2019}. The deviations are all within $20\textmd{ MeV}$. The systematic error of the modified model for the heavy meson mass is about $20\textmd{ MeV}$.

\begin{center}
 
\tabcaption{\label{tab:heavymesons} Masses of ground state heavy scalar and axial-vector mesons (in GeV). $M^{\textmd{DSE}}_{f\bar{g}}$ is the Dyson-Schwinger equation results of mesons with quark flavor content $f\bar{g}$. The first error is from the extrapolation, and the second is from varying the parameters. $M^{\textmd{expt.}}_{f\bar{g}}$ is the corresponding experiment results \cite{Zyla2020}. The underlined values in this table and TABLE \ref{tab:heavylightmesons} are used to fit the parameters in Eq. (\ref{eq:gluondInfrared}). $\bar{M}^{\textmd{RL}}_{c\bar{b}}$ are the results by estimating the errors from the original RL approximation calculation carefully \cite{Chen2020}. $M^{\textmd{LQCD}}_{c\bar{b}}$ are the lattice QCD results \cite{Mathur2018}. $M^{\textmd{QM}}_{c\bar{b}}$ are the quark model results \cite{Li2019}.}
\begin{tabular}{c|c|c|c}
\hline
$J^{\textmd{PC}}$ & $0^{++}$ & $1^{+-}$ & $1^{++}$ \\
\hline
$M^{\textmd{DSE}}_{c\bar{c}}$ & 3.415(0)(0) & 3.502(4)(10) & 3.507(4)(5) \\
$M^{\textmd{expt.}}_{c\bar{c}}$ & \underline{3.415} &  3.525 & 3.511 \\
\hline
$M^{\textmd{DSE}}_{b\bar{b}}$ &  9.859(0)(0) & 9.905(1)(3) & 9.897(1)(2) \\
$M^{\textmd{expt.}}_{b\bar{b}}$ & \underline{9.859} & 9.899 & 9.893 \\
\hline\hline
$J^{\textmd{P}}$ & $0^{+}$ & $1^{+}_1$ & $1^{+}_2$ \\
\hline
$M^{\textmd{DSE}}_{c\bar{b}}$ & 6.700(1)(4) & 6.743(2)(3) & 6.780(3)(3) \\
$\bar{M}^{\textmd{RL}}_{c\bar{b}}$ & 6703 &  6745 & 6781 \\
$M^{\textmd{LQCD}}_{c\bar{b}}$ & 6712 &  6736 & -- \\
$M^{\textmd{QM}}_{c\bar{b}}$ & 6714 &  6757 & 6776 \\
\hline
\end{tabular}
\end{center}

\subsection{The heavy-light mesons}

The results of the heavy-light meson masses are listed in table \ref{tab:heavylightmesons}. Except the scalar ($J^{P} = 0^{+}$) D meson mass ($M_{n\bar{c}}$) being used to fit the parameters, all the other masses are output quantities. Comparing the DSE results with the well established experiment value, i.e. the masses of the lowest axial-vector D meson ($M^{\textmd{expt.}}_{n\bar{c}}$), B meson ($M^{\textmd{expt.}}_{n\bar{b}}$) and $B_s$ meson ($M^{\textmd{expt.}}_{s\bar{b}}$), the systematic error of the modified model for the heavy-light mesons is conservatively estimated to be $40\textmd{ MeV}$. Now look at the fourth row of table \ref{tab:heavylightmesons}. The DSE results of the $D_s$ meson masses are unnaturally larger than the experiment value, from around $60\textmd{ MeV}$ to $120\textmd{ MeV}$. This supports that the ground state scalar and axial vector $D_s$ mesons may be not pure $q\bar{q}'$ states \cite{Chen2004,Faessler2007,Faessler2007a}. The other experiment values are still missing at present and my calculations make a prediction for them.

\begin{center}
 
\tabcaption{\label{tab:heavylightmesons} Masses of ground state heavy-light scalar and axial-vector mesons (in GeV). The notations are the same as TABLE \ref{tab:heavymesons}. For isospin multiplets, the average value of the experiment data is cited. The experiment data with a $\dag$ on the shoulder indicate that they may not be pure $q\bar{q}'$ states \cite{Chen2004,Faessler2007,Faessler2007a}.}
\begin{tabular}{c|c|c|c}
\hline    
$J^{\textmd{P}}$ & $0^{+}$ & $1^{+}_1$ & $1^{+}_2$ \\
\hline
$M^{\textmd{DSE}}_{n\bar{c}}$ & 2.333(6)(0) & 2.443(8)(9) & 2.569(10)(21) \\
$M^{\textmd{expt.}}_{n\bar{c}}$ & \underline{2.333} &  2.421 & -- \\
\hline
$M^{\textmd{DSE}}_{n\bar{b}}$ &  5.636(13)(5) & 5.693(10)(2) & 5.836(10)(6) \\
$M^{\textmd{expt.}}_{n\bar{b}}$ & -- & 5.726 & -- \\
\hline
$M^{\textmd{DSE}}_{s\bar{c}}$ & 2.434(0)(10) & 2.517(6)(17) & 2.596(11)(18) \\
$M^{\textmd{expt.}}_{s\bar{c}}$ & $2.318^\dag$ & $2.460^\dag$ & $2.535^\dag$ \\
\hline
$M^{\textmd{DSE}}_{s\bar{b}}$ &  5.755(7)(8) & 5.792(9)(12) & 5.878(12)(13) \\
$M^{\textmd{expt.}}_{s\bar{b}}$ & -- & 5.829 & -- \\
\hline
\end{tabular}
\end{center}

\subsection{The light mesons}

\begin{center}
 
\tabcaption{\label{tab:lightmesons} Masses of ground state light scalar and axial-vector $q\bar{q}$-states (in GeV). The notations are the same as TABLE \ref{tab:heavymesons}. For isospin multiplets, the average value of the experiment data is cited. The experiment data with a $\dag$ on the shoulder indicate that they may not a single $q\bar{q}'$ pole. The experiment value for the light scalar and isosinglet axial-vector meson masses are leaved empty, see the text.}
\begin{tabular}{c|c|c|c}
\hline
$J^{\textmd{PC}}$ & $0^{++}$ & $1^{+-}$ & $1^{++}$ \\
\hline
$M^{\textmd{DSE}}_{n\bar{n}}$ & 1.098(6)(8) & 1.220(10)(40) & 1.252(13)(31) \\
$M^{\textmd{expt.}}_{n\bar{n}}$ & -- &  1.230 & 1.230 \\
\hline
$M^{\textmd{DSE}}_{s\bar{s}}$ &  1.442(1)(57) & 1.533(5)(76) & 1.558(6)(62) \\
$M^{\textmd{expt.}}_{s\bar{s}}$ & -- & -- & -- \\
\hline\hline
$J^{\textmd{P}}$ & $0^{+}$ & $1^{+}_1$ & $1^{+}_2$ \\
\hline
$M^{\textmd{DSE}}_{n\bar{s}}$ & 1.331(4)(51) & 1.433(11)(68) & 1.502(11)(79) \\
$M^{\textmd{expt.}}_{n\bar{s}}$ & -- & $1.253^\dag$ & $1.403^\dag$ \\
\hline
\end{tabular}
\end{center}

The light scalar and axial-vector mesons are much more illusive. The $a_0(980)$ and $f_0(980)$ may be $K\bar{K}$ molecule states due to the color hyperfine interaction, and the actual $q\bar{q}'$ scalar mesons may become too light to be detected due to the same reason \cite{Godfrey1985}. For the isosinglet light axial-vector mesons, flavor mixing should be taken into account \cite{Chen2015}. It's beyond the scope of this paper to explore those questions. However, we can still draw some conclusions from present DSE results. All the DSE results for the light mesons are the output quantities, and are listed in table \ref{tab:lightmesons}. The masses of the lightest two axial vector mesons, i.e. $M^{\textmd{DSE}}_{n\bar{n}}$ with $J^{\textmd{PC}} = 1^{++}$ and $1^{+-}$ are consistent with the experiment values, the masses of $a_1$ and $b_1$. So Eq. (\ref{eq:gluonfdInfrared}) and (\ref{eq:gluondInfrared}) is still effective for the light mesons.

Note that the parameter dependency of the light meson masses is much larger than those of heavy and heavy-light mesons. The overall systematic error of the light meson masses is also expected larger. However, the mass splittings of the two lowest axial-vector mesons of the same content should be reliable. Then look at the K mesons. From the DSE results, $M^{\textmd{DSE}}_{n\bar{s}}(1^{+}_2) - M^{\textmd{DSE}}_{n\bar{s}}(1^{+}_1) \approx 70 \textmd{ MeV}$. The experiment value $M^{\textmd{expt.}}_{n\bar{s}}(1^{+}_2) - M^{\textmd{expt.}}_{n\bar{s}}(1^{+}_1) \approx 150 \textmd{ MeV}$ seems unnaturally large. Some studies indicate that the experiment measured $K(1)(1270)$ may be actually two poles \cite{Geng2007,Meissner2020,Wang2020}. This is a possible reason of the deviation herein.

\section{Summary}\label{sec:summary}
\vspace{0.1cm}

In this work, I studied the spectrum of the scalar mesons and axial-vector mesons via Dyson-Schwinger equation and Bethe-Salpeter equation approach in the rainbow-ladder approximation. The interaction model is modified, i.e. adding a repulsive term to the original one that are effective for the pseudoscalar and vector mesons. The added term is flavor dependent and has 3 parameters. With the three parameters fitted by the masses of scalar ($J^{PC} = 0^{++}$) charmonium ($M_{c\bar{c}}$), bottomium ($M_{b\bar{b}}$) and the scalar ($J^{P} = 0^{+}$) D meson ($M_{n\bar{c}}$), the other outputted masses are all consistent with those experiment value that are well established. So this modified model is effective for all the heavy, heavy-light and light scalar and axial-vector mesons. This is the first time a systematic study of the spectrum of the scalar mesons and axial-vector mesons in the DSE approach is achieved. The systematic errors of the DSE results are estimated carefully. Comparing the DSE results with the experiment value, we find that the scalar and axial-vector $D_s$ mesons may not be pure $q\bar{q}'$ states and the experiment measured $K(1)(1270)$ may actually be two poles.

\acknowledgments{Acknowledgments: thanks Professor Xianhui Zhong for usefull discussion.}


\section{Appendix A}\label{sec:appendixA}

The three groups of parameters correspond to $\omega_u = 0.45, 0.50, 0.55 \textmd{ GeV}$ are listed in Tab.\ref{tab:parameters}.
The quark mass $\bar{m}_f^{\zeta}$ is defined by
\begin{eqnarray}
 \bar{m}_f^{\zeta} &=& \hat{m}_f/\left[\frac{1}{2}\textmd{Ln}\frac{\zeta^2}{\Lambda^2_{\textmd{QCD}}}\right]^{\gamma_m},\\
 \hat{m}_f	 &=& \lim_{p^2 \to \infty}\left[\frac{1}{2}\textmd{Ln}\frac{p^2}{\Lambda^2_{\textmd{QCD}}}\right]^{\gamma_m} M_f(p^2),
\end{eqnarray}
with $\zeta$ the renormalisation scale, $\hat{m}_f$ the renormalisation-group invariant current-quark mass and $M_f(p^2)$ the quark mass function.

\end{multicols}

\begin{center}
\tabcaption{\label{tab:parameters} Fitted parameters correspond to $\omega_{u/d} = 0.45, 0.50, 0.55 \textmd{ GeV}$. $\bar{m}_f^{\zeta=2\textmd{GeV}}$, $\omega_f$ and $D_f$ are measured in GeV. $\alpha$ and $\beta$ are of unit 1. $\delta$ is measured in $\textmd{GeV}^{\alpha+\beta-2}$.}
\begin{tabular}{c|c|c|c|c|c|c|c|c|c|c}
\hline \hline
\multicolumn{2}{c|}{} & \hspace*{-0.1cm} & \multicolumn{2}{|c|}{ Parameter-1} & \hspace*{-0.1cm} & \multicolumn{2}{|c|}{Parameter-2} & \hspace*{-0.1cm} & \multicolumn{2}{|c}{Parameter-3}\\
\hline
flavor& $\bar{m}_f^{\zeta=2\textmd{GeV}}$&\hspace*{-0.1cm}&\; $w_f $\; &\; $D_f^2$ \;&\hspace*{-0.1cm}&\; $w_f $\; &\; $D_f^2$ \;	&\hspace*{-0.1cm}	&\; $w_f $\; &\; $D_f^2$ \;\\ [0.5mm]
\hline
$u/d$	& 0.0049&\hspace*{-0.1cm}	& 0.450 & 1.133 		&\hspace*{-0.1cm}	& 0.500 & 1.060 		&\hspace*{-0.1cm}	& 0.550 & 1.014 \\
$s$	& 0.112	&\hspace*{-0.1cm}	& 0.490 & 1.090 		&\hspace*{-0.1cm}	& 0.530 & 1.040 		&\hspace*{-0.1cm}	& 0.570 & 0.998 \\
$c$	& 1.17	&\hspace*{-0.1cm}	& 0.690 & 0.645 		&\hspace*{-0.1cm}	& 0.730 & 0.599 		&\hspace*{-0.1cm}	& 0.760 & 0.570 \\
$b$	& 4.97	&\hspace*{-0.1cm}	& 0.722 & 0.258 		&\hspace*{-0.1cm}	& 0.766 & 0.241 		&\hspace*{-0.1cm}	& 0.792 & 0.231 \\
\hline
\multicolumn{2}{c|}{$\delta$} & \hspace*{-0.1cm} & \multicolumn{2}{|c|}{0.02385 } & \hspace*{-0.1cm} & \multicolumn{2}{|c|}{0.02210} & \hspace*{-0.1cm} & \multicolumn{2}{|c}{0.01959}\\
\multicolumn{2}{c|}{$\alpha$} & \hspace*{-0.1cm} & \multicolumn{2}{|c|}{0.2691 } & \hspace*{-0.1cm} & \multicolumn{2}{|c|}{0.2530} & \hspace*{-0.1cm} & \multicolumn{2}{|c}{0.2902}\\
\multicolumn{2}{c|}{$\beta$} & \hspace*{-0.1cm} & \multicolumn{2}{|c|}{7.893 } & \hspace*{-0.1cm} & \multicolumn{2}{|c|}{8.632} & \hspace*{-0.1cm} & \multicolumn{2}{|c}{9.826}\\
\hline \hline
\end{tabular}
\end{center}

\begin{multicols}{2}

\vspace{2mm}
\centerline{\rule{80mm}{0.1pt}}
\vspace{2mm}

%
%
%
%
%

\bibliography{PWave}

\end{multicols}

\clearpage

\end{document}